\documentclass[review]{elsarticle}

\usepackage{hyperref}

\usepackage{graphicx}
\usepackage{xcolor}
\usepackage[margin=1.0in]{geometry}

\usepackage{xurl}
\usepackage{amsthm}
\usepackage{amsmath}
\usepackage{bbm}

\usepackage{array}

\usepackage{wrapfig}
\usepackage{setspace}
\usepackage[normalem]{ulem}

\usepackage{natbib}
\setcitestyle{round, comma, sort&compress, super}

\journal{Journal of Biomechanics}

\biboptions{authoryear}

\nocite{*}

\begin{document}

\begin{frontmatter}

\title{A new flow-based design for double-lumen needles}

\author{Radu Cimpeanu$^{*}$}
\address{Mathematics Institute, University of Warwick, Coventry, CV4 7AL, United Kingdom \\ Mathematical Institute, University of Oxford, Oxford, OX2 6GG, United Kingdom}
\fntext[myfootnote]{$^{,*}$ Corresponding author: \url{radu.cimpeanu@warwick.ac.uk}}

\author{Alfonso~A.~Castrej\'{o}n-Pita}
\address{Department of Engineering Science, University of Oxford, Oxford, OX1 3PJ, United Kingdom}

\author{Lee Nai Lim}
\address{Gynaecology Department, Oxford University Hospital NHS Foundation Trust, Oxford, OX3 9DU, United Kingdom}

\author{Manu Vatish}
\address{Nuffield Department of Women’s Health and Reproductive Research, University of Oxford, Oxford, OX3 9DU, United Kingdom}

\author{Ektoras X. Georgiou}
\address{Gynaecology Department, Oxford University Hospital NHS Foundation Trust, Oxford, OX3 9DU, United Kingdom}

\begin{abstract}
Oocyte retrieval forms a crucial part of in vitro fertilisation treatment and its ultimate outcome. Standard double-lumen needles, which include a sequence of aspiration and flushing steps, are characterised by a similar success rate to single-lumen needles, despite their increased cost. A novel hydrodynamics-based needle called the OxIVF needle is proposed here, which is geared towards the generation of an internal flow field within the full follicular volume via laterally, rather than frontally, oriented flushing, leading to successful retrievals with no additional stress on the oocyte. A two-dimensional digital twin of the follicular environment is created and tested via multi-phase flow direct numerical simulation. Oocyte initial location within the follicle is varied, while quantities of interest such as velocity magnitude and vorticity are measured with a high level of precision. This provides insight into the overall fluid motion, as well as the trajectory and stresses experienced by the oocyte. A comparative benchmark set of tests indicated a higher success rate of the OxIVF needle of up to $100\%$, marking a significant improvement over the traditional double-lumen design whose success rate of no more than $75\%$ was also highly dependent on the location of the needle tip inside the follicle. All forces measured during these tests showcase how the oocyte experiences stresses which are no larger than at the aspiration point, with the flow field providing a gentle steering effect towards the extraction region. Finally, the flow generation strategy maximises oocyte yield, unlocking new capabilities in both human and veterinary contexts.

\end{abstract}


\end{frontmatter}

\section{Introduction}
\label{sec:intro}

Modern in vitro fertilisation (IVF) practice involves the use of exogenous hormones to stimulate multi-follicular development in the ovary. Briefly, within the cortex of the human ovary lie spherical structures termed follicles, surrounded by ovarian stroma --- a matrix of connective tissue and interstitial cells. Within each follicle lies an oocyte, surrounded by follicular fluid and supporting cells, including granulosa cells that surround the oocyte itself; forming the cumulus-oocyte complex (COC) \citep{taylor2019speroff}.

Oocyte retrieval, and more widely COC retrieval, is a key step in the IVF process and is undertaken when the follicular cohort reaches maturity. The procedure is typically performed under sedation using transvaginal ultrasound guidance, during which a needle is introduced into each follicle and the contents aspirated with the aim of recovering each oocyte \citep{eshre2019recommendations}. Approximately 60-80\% of follicles are seen to yield an oocyte \citep{girsh2021textbook}.

Oocyte number, embryo quality and the live birth rate have been shown to be closely linked \citep{toftager2017cumulative,vermey2019there}. For this reason, maximising the number of oocytes recovered is crucial to improving IVF outcomes. The concept of follicular flushing was introduced with the aim of improving oocyte yield. A dual-channel needle is used to carry out this intervention; the needle is comprised of two concentrically arranged channels: an inner aspiration channel and an outer flushing channel. Practically, following initial follicle aspiration, flushing involves introducing approximately 1-2 mL of culture media into the follicle and subsequently re-aspirating it. Clinical practice varies with regards to how exactly flushing is performed, with, for example, some practitioners carrying out one flushing cycle \citep{calabre2020follicular}, and others up to three cycles \citep{de2021evaluation}. 

Although follicular flushing may appear as a biologically plausible intervention, a recent Cochrane systematic review including more than 1500 women demonstrated no improvement to oocyte yield or the live birth rate with this intervention as compared to aspiration alone \citep{georgiou2022follicular}. Furthermore, based on the same source, there was evidence that flushing takes significantly longer to perform compared to aspiration alone. This may have repercussions for the patient, e.g. higher doses of sedative drugs, and the IVF clinic, e.g. suboptimal operating theatre use. 

It should be noted that all current dual-channel needle products on the market introduce flushing media in a direction that is parallel to the aspiration channel. Further, due to its small size (up to approximately $200$ $\mu$m), the oocyte along with the surrounding cumulus are not visible under ultrasound. We hypothesised that current market products do not access all parts of the follicle, hence ‘missing’ the opportunity to retrieve all oocytes and therefore accounting for the recovery rates quoted above. We then went on to hypothesise that a novel needle, described herein, which allows for improved access of the flushing fluid to the follicle, would improve oocyte yield. 

While modelling based on computational fluid dynamics (CFD) approaches is extensively used in biomechanics, the IVF space has received relatively little attention in the context of this broad methodology. This is due to significant complexities in this area, among which we highlight ethical barriers in carrying out research in the field of human reproduction as particularly notable, as well as cross-disciplinary barriers. This investigative route has nevertheless been identified recently as a promising avenue for exploration \citep{batyuk2022biomechanical} given recent advances in terms of capabilities and computing power. A notable exception is the comparatively well studied embryo transfer process \citep{grygoruk2011fluid,yaniv2003biofluid}. Needle design has however been aided by computational tools in related areas, such as root canal irrigation \citep{boutsioukis2010evaluation}, while more broadly the usage of flow field knowledge to assist medical procedures has innovated design in areas such as kidney stone removal \citep{williams2020cavity}.

\begin{figure}[h]
\centering%
\includegraphics[width=0.99\textwidth]{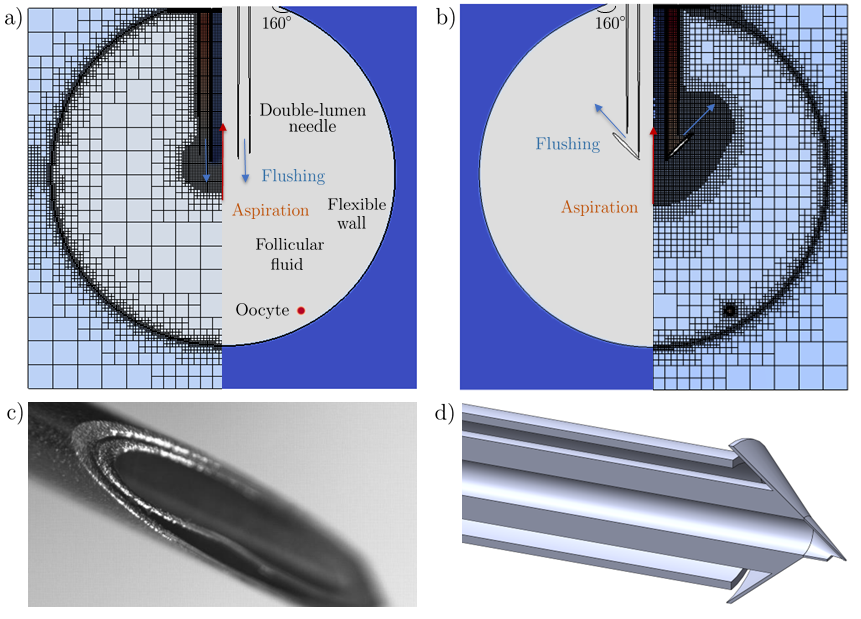}%
\caption{a) Schematic of the computational domain, containing the follicle environment (outer and inner fluid, flexible wall), the oocyte, as well as a typical needle insertion including aspiration and flushing regions (right), and discretisation of the in silico model allowing the visualisation of the adaptive grid cell refinement around features of interest (left). The imposed $160^{\circ}$ contact angle condition at the top boundary is also highlighted in both subplots. b) Proposed needle design showcasing lateral flushing functionality, with the different fluid regions (left) and discretisation (right) highlighted as part of the same simulation snapshot. c) Image of standard double-lumen needle as counterpart to the schematic in panel a), highlighting the suction and aspiration regions. d) Rendering of new needle design shown in panel b), with flow redirectioning in view of the additional geometrical features.}%
\label{fig:figure1}
\end{figure}

In the present work, we have made use of state-of-the-art computational capabilities in order to navigate this problem from a completely new perspective. An accurate in-silico model of the follicular environment (see Figure~\ref{fig:figure1}a) has been developed and tested via direct numerical simulation. This ensures a high degree of accuracy within a complex flow environment that allows visualisation of quantities of interest, from flow properties to the trajectory and forces experienced by the oocytes. This detailed information provided us with new insight into the vortical structures inside the flow, which can be used in order to vastly increase the efficiency of the needle design. More concretely, knowledge of the vorticity field coupled with the modelled oocyte motion has allowed us to propose a laterally oriented flushing mechanism (illustrated in Figure~\ref{fig:figure1}d)) which uses the flow field itself in order to steer the oocytes towards the aspiration areas, while preventing crushing the biological matter into the side of the needle wall. 

In what follows, we describe the mathematical modelling and computational platform as part of Section~\ref{sec:methods}, while also providing details on the parameters and protocols used. A systematic characterisation of the flow behaviour and its effects on the oocyte yield is provided in Section~\ref{sec:results}, while conclusions and recommendations are highlighted as part of Section~\ref{sec:discussion}.

\section{Materials and methods}
\label{sec:methods}

We have constructed the numerical infrastructure for this study using the open-source package Gerris \citep{Popinet_2003, Popinet_2009}, with subsequent development based on its newer version Basilisk \citep{popinet2015quadtree, Basilisk_Website}. Both finite volume based partial differential equation solvers, they have underpinned progress in the community for more than two decades. Important highlights in terms of capabilities are given by the well-known volume-of-fluid (VOF) method \citep{hirt1981volume} to treat interfaces between different fluid regions in an accurate manner, while the quadtree/octree-based adaptive mesh refinement capabilities and parallelisation features make it an excellent resource for multi-scale multi-physics problem implementation. The infrastructure relies on a staggered in time discretisation of the volume fraction and pressure, which leads to a Poisson equation for the pressure field via a time-splitting projection method \citep{chorin1969convergence} as the most intensive element of the calculation. This procedure results in a multilevel method which is second-order accurate in both space and time, with additional problem-dependent criteria (from geometry to dynamics) guiding an efficient adaptive mesh refinement process.
The accurate and robust approaches for fundamental fluid dynamics phenomena, coupled with the ability to develop and specialise the system towards the geometry and features of interest here, have resulted in a versatile implementation that has allowed us to test the new needle in a space in which flow effects can be readily monitored and embedded in the design workflow. 

The underlying governing equations for the modelled flow problem are the full two-dimensional Navier-Stokes equations, in which we have made several assumptions in order to make the problem tractable and focus on some of the fundamental features of our setup. In particular, the dimensionality reduction preserves the key elements of the dynamics we are trying to capture while significantly lowering the computational cost. We have assumed a three-fluid system, with an outer viscous region, a follicular fluid acting as the primary environment of interest, and the oocyte itself. Note that three-phase flow implementations have only been recently addressed rigorously in the fluid dynamics space, with progress encouraged by efforts in related interfacial flow areas such as drop coalescence and impact \citep{fudge2021dipping, fudge2023splashing,marcotte2019ejecta}. All fluids have been modelled as Newtonian, with effective physical properties aligned with the biological features of interest. In particular, by denoting the three fluid regions in question with subscripts $(\cdot)_i$, and $i=1,2,3$ referring to the oocyte, follicular, and external regions, respectively (please consult Figure~\ref{fig:figure1}a), we set constant densities $\rho_1 = \rho_2 = 998$ kg/m$^3$, $\rho_3 = 1.5 \cdot 998$ kg/m$^3$ and constant dynamic viscosities $\mu_1 = \mu_2 = 0.002$ kg/ms, $\mu_3 = 0.02$ kg/ms. The external or outer fluid regions is thus modelled as having a larger density and viscosity than the follicular environment, which has been assumed to behave similarly to water. The oocyte also inherits the same properties for the interior fluid. A surface tension coefficient of $\sigma_{12} = 0.024$ kg/s$^2$ is prescribed between the oocyte and the follicular fluid, while a much stronger $\sigma_{23} = 0.15$ kg/s$^2$ is provided for the follicular wall. This is intended to replicate some of the deformability of this environment in a simplified form given the neglected biological effects that may effect its structure. While non-Newtonian features could also further enrich this setup, we decided to concentrate on the fundamental phenomena driving flow behaviour. Furthermore, given the relatively quick timescales and strong inertial effects in our physical system, gravitational effects have also been neglected herein. We thus have
\begin{align}
\rho_i(\partial_t\mathbf{u}_i+\mathbf{u}_i\cdot \nabla \mathbf{u}_i) &= -\nabla p_i + \nabla \cdot (2\mu_i \mathbf{D})+\sigma_{ij} \kappa \delta_s \mathbf{n}, \label{eq:navierstokes1}\\
\nabla \cdot \mathbf{u}_i &= 0,
\label{eq:navierstokes2}
\end{align}
where subscript $i$ is used to indicate each fluid region and its properties, 
$D_{kl}\equiv (\partial_ku_l+ \partial_lu_k)/2$ is the deformation stress tensor, $\rho$, $\mu$ and $\sigma$ the density, viscosity and surface tension respectively, $\kappa$ and $\mathbf{n}$ the curvature and normal vector of the interface respectively and $\delta_s$ the Dirac delta function that takes a value of $1$ at the interface and $0$ elsewhere such that the surface tension term only applies at the interface. With the above approximations, the system is closed by typical interfacial conditions (continuity of velocities, normal and tangential stress balances, and the kinematic condition), as well as boundary conditions describing relevant velocity outflow and inflow modelling aspiration and flushing, respectively, while other computational domain boundaries in the external region are set to be free-slip in order to avoid them affecting the flow inside the region of interest. The follicular wall has a prescribed angle of $160^{\circ}$ for the top boundary to ensure the approximately circular shape is preserved, but is otherwise free to move and change volume with respect to the aspiration and flushing protocols prescribed.

As we transition between each fluid region to another, volume-of-fluid methods make use of an additional variable to distinguish the phases. For a two-phase case this can be a single scalar variable which takes a value of $1$ in one of the phases and $0$ in the other and a value between the two in interfacial regions \citep{scardovelli1999direct}. This extra variable, also known as a colour function, then allows other properties (such as density or viscosity) to be expressed in terms of it. For example if we denote the colour function as $c$ and the values of a variable for phases $1$ and $2$ respectively, then we may express a physical property variable such as density as $\rho =c \rho_1+(1-c)\rho_2$. The colour function is also governed by a conservation equation, which is a requirement that in addition to conserving the overall mass of fluid, the mass of each individual phase must also be conserved, formulated as
\begin{equation}
\partial_tc+\nabla \cdot(c\mathbf{u})=0.
\label{eq:multiphasecons}
\end{equation}

The above setup naturally requires a generalisation when three rather than two fluids are considered, with two colour functions instead of one being utilised instead in order to differentiate between the fluid regions in a pair-wise manner, and prevent artificial coalescence or rupture. Therefore, now generalising the colour function $c$ definition into subcomponents $c_1$ (representing the oocyte) and $c_2$ (representing the intra-follicular environment), we expand our previous definition for the physical properties of interest such as density to $\rho = c_1 \rho_1 + c_2 \rho_2 + (1-c_1-c_2) \rho_3$ in order to characterise each component in our three-fluid system.

Parametric choices are dictated by known biological quantities in the follicular environment, which help steer choices in typical values. Thus the full computational domain lengthscale measures $0.02$ m, with the initial circular follicular wall occupying $90\%$ of it in each dimension. The radius of the oocyte itself is prescribed as $200\ \mu$m, based on a typical oocyte sizing of $\mathcal{O}(100)\ \mu$m, while accounting for some of the surrounding cell presence as part of the COC. We have selected a typical $16$-gauge needle sizing as proof of concept, with a $1.2$ mm diameter of the inner lumen, and a further $0.2$ thickness mm for the outer lumen channel. It is inserted into the follicle such that it reaches $8$ mm into the fluid region, and appears as an embedded solid inside the computational domain. The arrow-based feature of the OxIVF needle which ensures the redirection of the flow is attached concentrically to the inner lumen (as showcased in Figure~\ref{fig:figure1}b/d).

The velocity scale is given by the driving aspiration mechanism. Using known information from the literature \citep{eshre2019recommendations}, we know that pressures used in practice lie between $100$ and $200$ mmHg, with lack of data as to potential damage of higher pressures on oocytes. Under this guideline, we can make use of a Hagen-Poiseuille flow solution to map pressure settings with the volumetric flow rate, a quantity which is easy to monitor and control in practice via $\textstyle{\Delta p = 8 \mu Q L / (\pi R^4)}$, where both needle length $L$ and radius $R$ being adapted to this setup. The real life longitudinal lengthscale of the needle of $\mathcal{O}(10)$ cm is sufficiently large to justify this assumption of the flow profile as having been sufficiently well established by this stage, which only requires an $O(1)$ mm lengthscale for our typical flow conditions. This has led us to an estimate of the average flow profile velocity of $0.2$ m/s, with a parabolic velocity profile prescribed as a boundary condition in the aspiration area. The flushing velocity is applied similarly but with an opposite sign and adapted value in the outer lumen regions.

The oocyte initial locations were designed such that
the interior follicular space is covered uniformly with a tractable number of numerical experiments, including regions behind the tip of the needle. Six approximately equiangular locations, starting from the south pole of the follicular volume and moving upwards in the near vicinity of the well, have been prescribed along the contour of a concentric circle of radius $0.4$ relative to the $0.45$ initial radius of the follicle. The placement near the wall has been considered in order to replicate the in vivo environment, whereby the oocyte is often found in the vicinity of the follicular wall. 

The non-dimensionalisation procedure uses the full domain size as reference lengthscale and the aspiration velocity as reference velocity scale, thus setting the relevant timescale. The follicular fluid was nominated as the reference liquid, with all other physical quantities normalised with respect to it. Finally, pressure was nondimensionalised on an inertial scaling via $p \sim \rho_l V^2$. As implemented above, the simulations were executed over $30$ dimensionless time units, which translates to $3$ s in dimensional time. Aspiration and flushing were set to occur at the same time for the first second, after which only aspiration took place. As previously noted, follicular flushing has not resulted in improvement of retrieval rates and the process itself often leads to procedures with a longer duration, higher levels of drug use for the patients and poorer use of clinical resources \citep{mok2013follicular, von2017randomized}, which is why its use needs to be both justified and well optimised. Thus, rather than the conventional aspirate-flush-aspirate procedure, we have opted for a concurrent aspiration and flushing for a duration of $1$ s which was sufficient for the flow field movement to fully develop and act on the oocyte, followed by a single aspiration segment that either ensures successful retrieval or results in the oocyte being trapped in a region often to the side of the needle as the follicular wall collapses. During the complex dynamics above, we store and process the location of the oocyte (via its interfacial shape and center of mass location), the interface shape of the follicular wall and its volume, as well as norms and extremal values of quantities of interest such as pressures, velocity components, and vorticity. Combined, these allow us to explore flow features and efficiency aspects in order to conduct a rigorous comparative performance analysis between a standard double-lumen needle and the newly proposed design, which is expanded upon in Section~\ref{sec:results}. Further details on the computational setup in view of refinement, parallelisation, execution time as well as validation are provided as part of~\ref{sec:app1}.

\section{Results}
\label{sec:results}

\begin{figure}[!h]
\centering%
\includegraphics[width=0.9\textwidth]{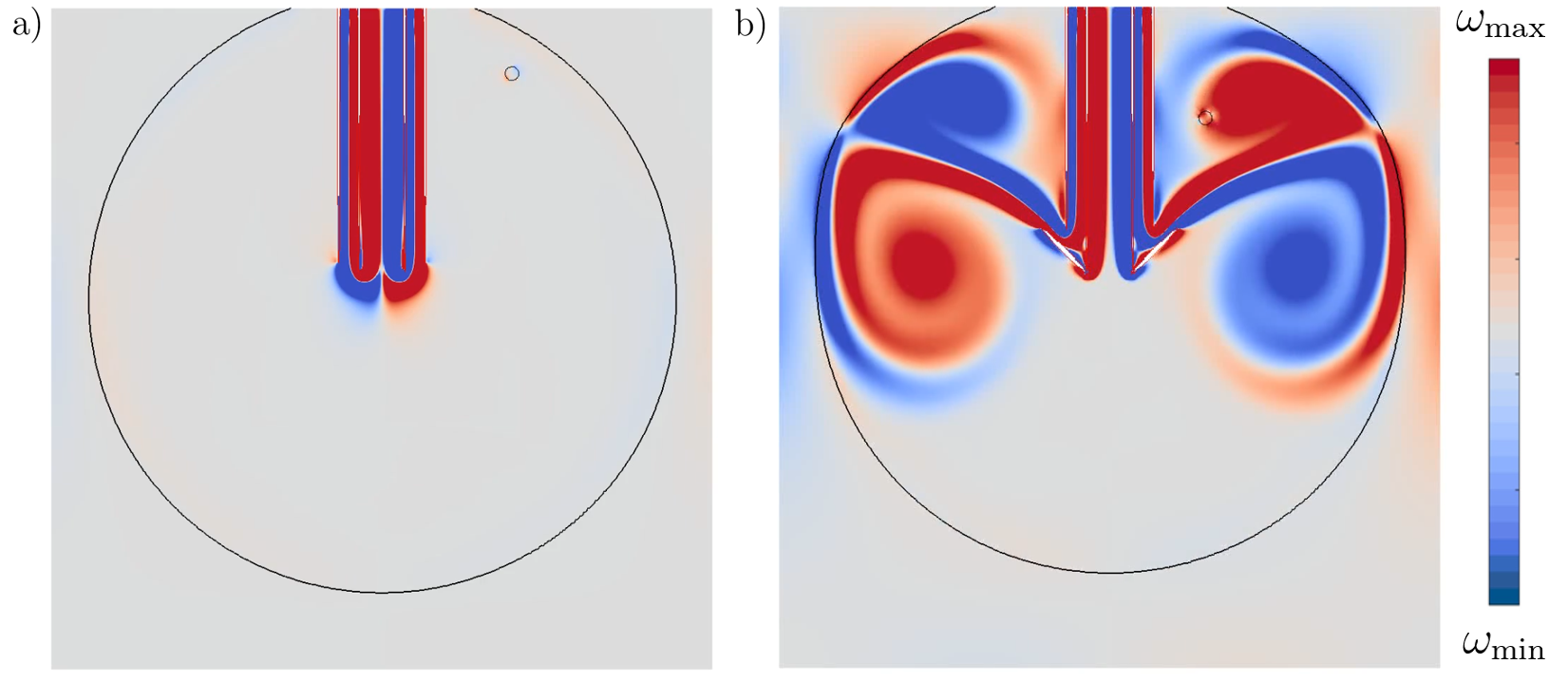}%
\caption{Magnitude of the vorticity field during the joint aspiration-flushing cycle at dimensionless $t=0.25$ for a) the conventional double-lumen needle and b) the OxIVF needle. The colourscale is adaptive and indicates the minimum vorticity value $\omega_\textrm{min}$ in blue and maximum vorticity value $\omega_\textrm{max}$ in red at each point in time. The oocyte (top right hand quadrant) and the deformation of the follicular wall under the action of the flow can also be observed.}%
\label{fig:figure2}
\end{figure}

The numerical framework constructed for the in silico experimentation in this context provides us with the opportunity to comprehensively investigate the induced flow field under the action of the flushing and/or aspiration dynamics imposed. We are particularly interested in examining the vortex dynamics inside the follicular volume as an indicator for the type of flow field the oocyte is subjected to during its extraction lifecycle. In Figure~\ref{fig:figure2} we present an early stage snapshot of the vorticity field magnitude $\omega$ for both the standard double lumen needle (panel a) and new design (panel b). This is taken at $t=0.25$ s, a quarter of the timescale during which aspiration and flushing are applied simultaneously.

For the standard double lumen needle it is clear that both its orientation and close proximity of the flushing and aspiration regions induce detrimental effects. Even with aspiration being turned off and flushing active, the flushing fluid is steered frontally, with the vast majority of the follicular volume hosting potential oocytes remaining motionless. When both aspirating and flushing are active, a significant part of the flow created by flushing is affected by the nearby aspiration. By contrast, the laterally imposed flushing of the new needle showcased in Figure~\ref{fig:figure2}b produces rich and constructive dynamics for the movement of the oocyte irrespective of its initial location. The direct flow is sufficiently strong to create counter-rotating vortex structures that occupy the full internal volume of the follicle which may deform its wall, but which are insufficiently strong to break it. This represents an ideal combination, with the oocyte being led towards the aspiration region which is also operational and can thus efficiently recover the biological matter. The clear separation between the aspiration and flushing regions means that flushing/aspiration protocols can be set much more robustly.

\begin{figure}[!h]
\centering%
\includegraphics[width=0.99\textwidth]{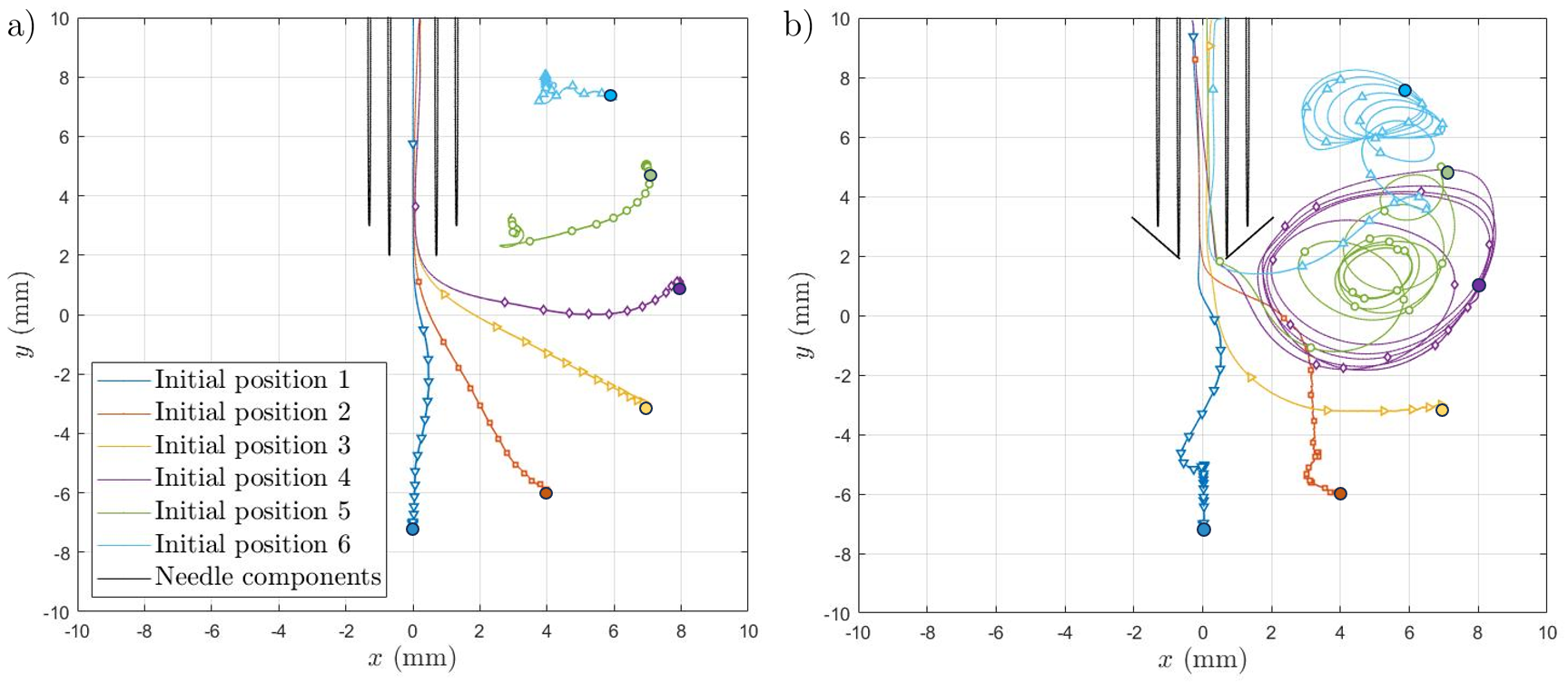}%
\caption{Oocyte trajectory timelines over the full duration of the simulation. Each initial position is encoded with a different colour during its evolution. For clarity, the initial locations of the oocytes for both datasets are highlighted using solid circles. a) Conventional double-lumen needle and b) OxIVF needle datasets. Oocytes are shown to either be successfully aspirated, or fail to reach the needle tip (double-lumen needle, initial positions 5 and 6) leading to failed attempts. }%
\label{fig:figure3}
\end{figure}

\begin{figure}[!ht]
\centering%
\includegraphics[width=0.99\textwidth]{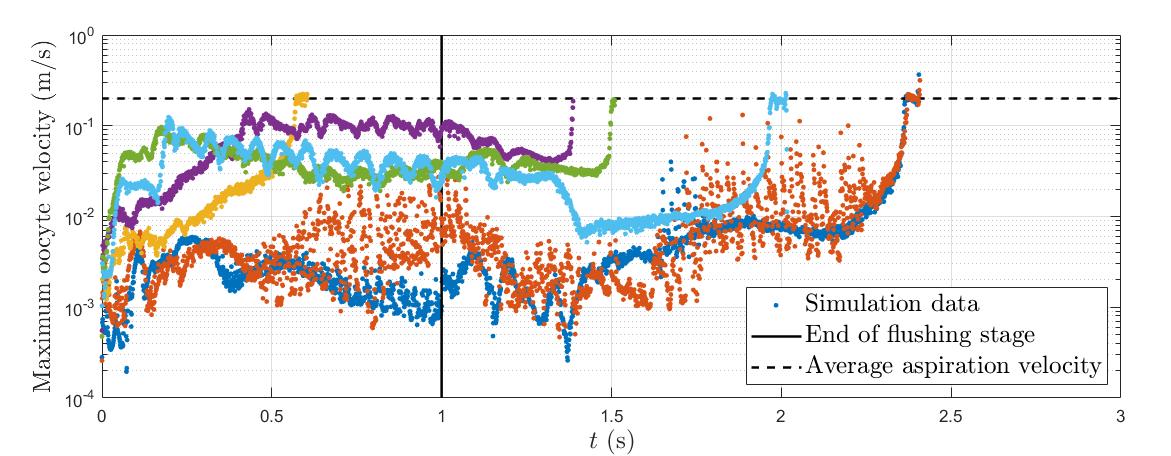}%
\caption{Maximum velocity experienced by the oocytes (measured over their respective areas) as a function of time using the OxIVF needle design. The vertical line indicates the point at wish the follicle volume-preserving flushing procedure is stopped and aspiration continues. The horizontal dashed line indicates the aspiration velocity prescribed at the outlet of the computational domain. The colour scheme from the previous Figure~\ref{fig:figure3} is retained, with the different colours indicating the different prescribed initial positions of the oocytes.}%
\label{fig:figure4}
\end{figure}

Figure~\ref{fig:figure3} provides an illustration of the trajectory of the center of mass of each oocyte (coloured differently in order to identify respective initial locations), with symbols plotted every $100$ iteration steps to provide insight into movement, which is further expanded upon quantitatively in Figure~\ref{fig:figure4}. The left hand side panel in Figure~\ref{fig:figure3}, detailing the standard double lumen needle performance panel, reveals successful outcomes for $66\%$ of the oocytes (four of the six numerical trials). Two test cases (in green and teal) have interrupted trajectories, which means that the follicular volume reduction encountered during the aspiration regime has led to the confinement of the oocytes in an unviable region (typically against the needle side wall). Additional trials with oocytes initially placed in the borderline region separating previous successful and unsuccessful aspiration outcomes (between the purple and green test cases), as well as in the challenging space behind the needle tip (in the region of the teal test case), suggest that the viable catchment area for the double-lumen needle measures approximately $75\%$ of the internal follicular space in this particular configuration. A more general proposed interpretation of these trajectories is that scenarios with initial oocyte locations placed in front of the needle lead to successful extraction once only aspiration is enabled, with flushing providing no concrete benefit, in agreement with conclusions from previously mentioned studies \citep{georgiou2022follicular} on the performance of standard double-lumen needles. While only several test cases are presented herein, further comprehensive numerical experimentation, including varying the needle tip location, strengthened the hypothesis that oocytes placed in the fluid volume frontal to the needle tip would be extracted successfully, with the number of unfeasible locations depending strongly on how deeply the standard double-lumen needle is inserted inside the follicle. Clinical practice is that the needle is placed in the centre of each follicle --- optimising this procedure is however difficult and subject to significant uncertainty, and is one of the challenges in operating the currently used standard double-lumen needle.

By contrast, in Figure~\ref{fig:figure3}b we illustrate the equivalent trajectories in the case of the new needle design. While oocytes located in the immediate front of the needle have similar behaviour, the three most challenging test cases are characterised by an ample motion induced by the counter-rotating vortex motion. This effect facilitates the movement of the biological matter from previously inaccessible challenging locations, such as behind the needle tip. While this additional movement may appear considerable, we note the alignment with the flow field and smoothness of the trajectories, which ensures gentle dynamics that we explore in more detail as part of Figure~\ref{fig:figure4}.

The level of stress experienced by the oocyte can also be carefully monitored in view of the underlying liquid formulation. This gives access to averaged metrics (e.g. over the surface area of its interface, or over its entire volume), as well as data on specific points, of which the center of mass of the oocyte represents a natural candidate for data examination. In Figure~\ref{fig:figure4} we provide information on the velocity experienced by the oocyte, presenting the maximum found within its volume. The spatial variation of this quantity is not pronounced given the lengthscale of the egg and its relatively rigid shell, however the maximum should nevertheless provide a more stringent quantitative interrogation of its dynamics. The same colour scheme as in the previous figure is used to identify each starting point. The velocity imposed as a boundary condition at the top of the computational domain, which we recall to be $0.2$ m/s, is also plotted for comparative purposes in a black dashed line. 

The most salient aspect in the evolution represented in Figure~\ref{fig:figure4} is that, despite the richness of the new vortical structure created by the new needle design and flushing procedure, at no point during the numerical experiment is the target velocity higher than that imposed as the aspiration velocity. This qualitatively translates to a comparatively mild dynamics, in which the vortical motion provides a gentle forcing from potentially unfavourable areas towards a viable aspiration region, which it finds in all cases. The oocytes are eventually directed towards the needle tip as a result of the follicular volume reducing, with the motion encouraged inside the follicle being enough to steer the oocyte towards the needle tip without it being isolated into a region where aspiration is no longer feasible, such as the side of the needle wall. In certain scenarios (see teal, green and purple datasets) periodic oscillations in the velocity pattern are also visible. These are indicative of the rotational motion induced by a flushing vortex, as shown in Figure~\ref{fig:figure3}b. It represents the mechanism through which the oocyte is set in motion and becomes a viable aspiration target from location in which a standard double lumen needle protocol would render it inert and ultimately unviable. Another quantity of interest is the shear stress experienced by the oocyte. Careful examination has indicated similar patterns of rich behaviour during the early stages of the flushing period for the new design leading to stresses of up to $0.1$ Pa found during the rotational motion. These are significantly lower than the stresses experienced by the oocytes once entry through the aspiration point and into the needle takes place, with local stress components typically ranging between $0.2$ and $0.4$ Pa. Through multiple examination routes we have thus established with confidence that the interior flow dynamics is the demonstrably non-invasive driving factor behind the successful operation of the new needle within the framework of our model.

\section{Discussion}
\label{sec:discussion}

In this study we have proposed and analysed a new theoretical framework for double lumen needle design. This represents a fundamental shift in how flow field information is utilised in this space, unlocking significant capabilities for improved design and optimisation. Laterally oriented flushing provided means towards accessing and utilising the full biological volume of interest, leading to a significantly improved outcome with no theoretical side effects in terms of stresses. The arrow-shaped geometrical feature that ensures the redirectionality of the flushing element acts as a versatile proof of concept element, with for example laser-cut insertions in the azimuthal direction representing a viable alternative that may prove even more usable in practice. While in vitro fertilisation procedures have provided the motivational foundation for this study, we envision that the computational modelling platform may become useful in other contexts, such as abscess drainage.

An important dimension of the new needle design is provided by its activation of the internal flow field inside the follicular volume irrespective of its initial placement. Thus the location of the oocyte relative to the needle tip would no longer represent a critical factor in its success rate in terms of successful retrieval. 

Several avenues for generalisation and improvement remain even in the two-dimensional case, the most notable of which we highlight below:
\begin{enumerate}
    \item The assumption of Newtonian fluid formulation is an important one, with rheological aspects and further biological complexity becoming exploratory avenues of interest.
    \item The formulation of interfacial conditions for both the follicular and oocyte walls has been designed to mimic some of the real-life flexibility observed, however is nevertheless simplified. A more complete fluid-structure interaction formalism, including elasticity effects, represent meaningful future directions.
    \item All of the above require significant numerical implementation efforts while retaining features of interest e.g. adaptive mesh refinement, parallelisation, many of which represent large scale cutting edge endeavours in the computational modelling space.
    \item Added variation in view of physical setup (follicular and oocyte sizes, needle gauge) as well as operational testing (different aspiration-flushing protocols) should support the above generalisations and provide a comprehensive assessment in terms of performance and application-specific potential. 
\end{enumerate}

\begin{figure}[h]
\centering%
\includegraphics[width=0.95\textwidth]{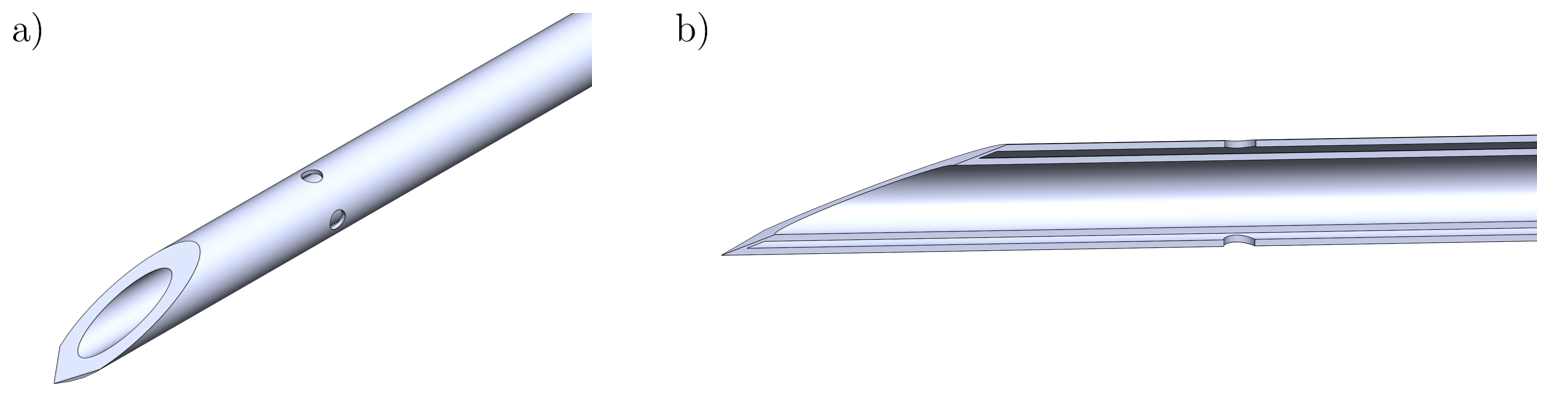}%
\caption{Alternative OxIVF needle design illustrating lateral orifices for the flushing fluid travelling through the outer lumen to be redirected outwards with the aim of achieving a similar intra-follicular flow excitation behaviour to that induced by the arrow-head OxIVF needle.
a) General three-dimensional view  and b) cross-section view of this design variation.}%
\label{fig:figure5}
\end{figure}

Bringing the computational model of the needle even closer to its real-life counterpart would significantly benefit from the consideration of a full three-dimensional structure. This introduces added complexity (in particular a need for a robust implementation of three-phase flows and embedded solid interaction in a volume-of-fluid context) and computational resource requirements, but would also provide valuable exploratory opportunities. If targeted correctly, it can aid early stages of the design pipelines and inform both structural and operational aspects of the manufacturing and usage processes. It would also allow the introduction of non-axisymmetric three-dimensional features that stretch beyond testing capabilities in two dimensions. To this end, an alternative OxIVF needle design is showcased in Figure~\ref{fig:figure5}. While retaining the same flow-enhancing functionality behind the previously discussed arrow-shaped feature, the orifices in this alternative design are likely to provide a less invasive insertion process and hence less traumatic patient experience, while retaining the lateral flushing elements that drive the added efficiency of the design. Early investigations using Lagrangian particle tracking have shown promise in terms of generating rich internal dynamics, with a detailed generalised study currently in progress.

Since this in silico work was undertaken, the proposed designs have been patented \citep{vatish2022needle, europeanpatent}. Furthermore, following manufacture of a prototype similar to the OxIVF needle design outlined in Figure~\ref{fig:figure5}, we have undertaken an ex vivo project using ovaries obtained from cattle, with veterinary trials also successfully confirming the benefits of employing this design \citep{simmons2022}, yielding real-world data demonstrating superior oocyte yield as compared to existing dual-channel needles. Following on from this work, we have conducted an in vivo cattle project with similarly encouraging results, which is in the final stages of preparation for further dissemination. We are also currently in the process of building on our animal data in other animal models, whilst also preparing to undertake human testing. 

As indicated by the latest Cochrane systematic review \citep{georgiou2022follicular}, follicular flushing does not improve IVF outcomes. Our study provides an insight as to why this may be the case, and encouragingly proposes a solution in the form of a new needle taking the detailed fluid mechanical processes into account. The approach, and level of actionable detail it provides, may be of benefit to a number of related challenges, such as drainage and/or aspiration of pus in peritonsilar abscesses \citep{Chang2016quincy}, as well as joint aspiration \citep{Hermans2011jointasp}.

\section*{CRediT authorship contribution statement}
\textbf{Radu Cimpeanu}: Conceptualization, Writing – original draft, Methodology, Software, Validation, Data curation, Visualization, Formal analysis.

\textbf{Alfonso~A.~Castrej\'{o}n-Pita}: Conceptualization, Writing - review \& editing, Methodology, Investigation, Resources.

\textbf{Lee Nai Lim}: Writing - review \& editing.

\textbf{Manu Vatish}: Conceptualization, Writing - review \& editing.

\textbf{Ektoras X. Georgiou}: Conceptualization, Writing - original draft, Supervision, Project Administration.

\section*{Conflict of competing interest} The authors declare that they have no known competing financial interests or personal relationships that could have appeared to influence the work reported in this paper.

\section*{Data availability statement} For the purpose of open access, the author has applied a Creative Commons Attribution (CC-BY) licence to any Author Accepted Manuscript version arising from this submission. Source code and datasets will be made available to interested users upon reasonable request.

\section*{Acknowledgements} 
 RC gratefully acknowledges the support of the Mathematical Institute at the University of Oxford through funding via the Hooke Research Fellowship and access to the departmental high performance computing facilities that supported the work reported here. All authors are thankful to Oxford University Innovation for funding obtained via the University Challenge Seed Fund supporting the most recent stages of this study, as well as helpful discussions in the early stages of the investigation. We also thank Dr. Benjamin D. Fudge (University of Oxford), who has provided support in generating some of the schematic models in Figures~\ref{fig:figure1} and \ref{fig:figure5} in this manuscript.  Finally, the authors would like to thank the anonymous referees whose comments helped improve upon a previous version of this manuscript.

\appendix
\section{Computational details and validation}
\label{sec:app1}

The adaptive mesh refinement (AMR) strategy used as part of the direct numerical simulation platform is given by a high degree or resolution allocated to interfacial and geometry (needle) wall locations as primary criteria, with changes in magnitude of velocity components, as well as vorticity as secondary criteria.
The minimum grid cell size (corresponding to the highest resolution) was $40\ \mu$m equivalent to $2^9 =512$ grid cells per dimension. The mesh is adapted at every timestep, with prescribed tolerances (of $\mathcal{O}(10^{-4})$ for the needle, oocyte and follicular wall locations, and of $\mathcal{O}(10^{-2})$ for changes in the magnitude of the velocity components and the vorticity) dictating an appropriate refinement level between $2^4$ and $2^9$ cells per dimension, which we call resolution levels $4$ and $9$, respectively. This naturally leads to more resources being allocated in key regions of the flow, as highlighted in Figure~\ref{fig:figure1}. While a uniform grid with this setting would become computationally prohibitive, with AMR the effort was a much more tractable typical value of $50000$ gridcells (between $20000$ and $100000$ depending on parametric values), with a simulation runtime of $500$ CPU hours executed over $4$ or $8$ cores on high performance computing clusters.

\begin{figure}[!h]
\centering%
\includegraphics[width=0.99\textwidth]{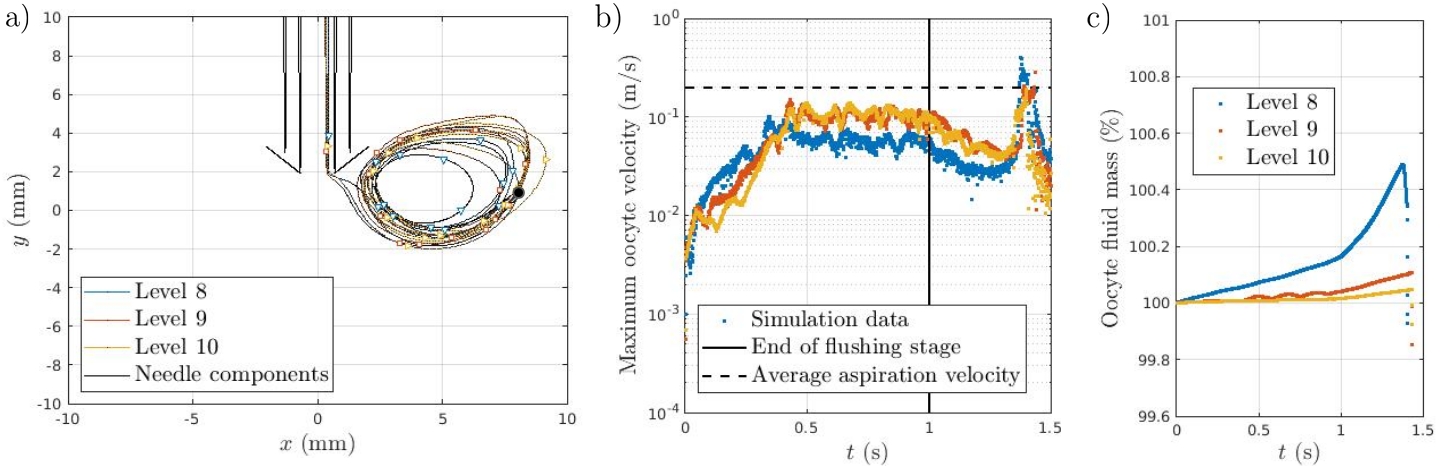}%
\caption{Resolution study using an adaptive mesh refinement setup with minimal computational cell size varying from $1/2^8$ (level $8$) to $1/2^{10}$ (level $10$) relative to the domain size, for a representative test case under the action of the OxIVF needle. a) Oocyte trajectory timeline over the full duration of the simulation for each level. The initial oocyte location is marked with a black circle. b) Maximum velocity experienced by the oocytes (measured over their respective areas) as a function of time. c) Oocyte mass conservation as a function of time.}%
\label{fig:figure6}
\end{figure}

Oocyte and follicular volumes, as well as velocity and vorticity-based norms, were monitored as a function of time in order to ascertain the mesh independence of the calculations to an acceptable tolerance level, which was deemed to be $0.1\%$ in view of mass conservation and $0.01$ (non-dimensionally) in terms of the other normalised quantities examined. Figure~\ref{fig:figure6} provides a summary viewpoint of a typical resolution study performed in order to assert the robustness of the setup, restricted for brevity to a subset of the quantities of interest. Early numerical experimentation had revealed that a minimum cell size of $2^7$ cells per dimension (or level $7$) is insufficient to accurately describe the geometrical details of the needle. We thus restricted our comparative investigation to levels $8$-$10$. Referring to panel a), while the oocyte trajectories are qualitatively similar and the overall agreement is already encouraging, quantitative details and in particular mass conservation properties in panel c) for level $8$ calculations lead us to conclude that we would benefit from an accuracy level provided by an AMR setup with a minimum cell size given by level $9$. Setting maximum resolution levels of either $9$ or $10$ leads to results that are in very good agreement both qualitatively and quantitatively, abiding by the desired $0.1\%$ tolerance criterion for mass conservation, and thus providing an overall solid foundation for the parametric studies described in Section~\ref{sec:results}. Guided by similar insight obtained from other test cases and in view of efficient resource usage, we have thus opted for a setup with a maximum refinement level of $9$ for our numerical campaign.



\bibliography{OxIVF_Main_Accepted}

\end{document}